\documentclass[aps,prd,preprintnumbers,nofootinbib,showpacs,onecolumn]{revtex4}%
\usepackage[dvips]{graphicx,psfrag}
\usepackage{color} 
\usepackage{amsmath,braket}
\usepackage{amssymb}
\usepackage{bm}
\newcommand{\bea}{\begin{eqnarray}}
\newcommand{\eea}{\end{eqnarray}}
\newcommand{\be}{\begin{equation}}
\newcommand{\ee}{\end{equation}}

 \makeatletter  
\def\alt{\mathrel{\mathpalette\gl@align<}}
\def\agt{\mathrel{\mathpalette\gl@align>}}
\def\gl@align#1#2{ \lower.6ex\vbox{\baselineskip\z@skip\lineskip\z@
\ialign{ $\m@th#1\hfil##\hfil$\crcr#2\crcr\sim\crcr }} } \makeatother

\begin{document}

%

\title{
Quantum radiation from a particle in an accelerated motion 
coupled to\\ vacuum fluctuations
}
\vspace{1cm}

%
\author{
Naritaka Oshita${}^{1,2}$, Kazuhiro Yamamoto${}^{3,4}$, and Sen Zhang${}^5$
}
 \vspace{.5cm}

\affiliation{
$^{1}$Department of Physics, Graduate School of Science, The University of Tokyo, Bunkyo-ku, Tokyo 113-0033, Japan \\
$^{2}$Research Center for the Early Universe (RESCEU),Graduate School of Science, The University of Tokyo, Bunkyo-ku, Tokyo 113-0033, Japan\\
$^{3}$Department of Physical Science, Graduate School of Science, Hiroshima University,
         Higashi-Hiroshima 739-8526, Japan \\
$^{4}$Hiroshima Astrophysical Science Center, Hiroshima University,
         Higashi-Hiroshima 739-8526, Japan \\
$^{5}$Okayama Institute for Quantum Physics,
Kyoyama 1-9-1, Kita-ku, Okayama 700-0015, Japan
}

\vspace{2cm} 
\begin{abstract}
A particle in a uniformly accelerated motion exhibits Brownian random 
motions around the classical trajectory due to the coupling to the 
field vacuum fluctuations. Previous works show that the Brownian random 
motions satisfy the energy equipartition relation. This thermal property 
is understood as the consequence of the Unruh effect.  
In the present work, we investigate the radiation from the thermal 
random motions of an accelerated particle coupled to vacuum fluctuations.
The energy flux of this radiation is negative of the order smaller than the 
classical radiation by the factor $a/m$, where $a$ is the acceleration 
constant and $m$ is the mass of a particle.
The results could be understood as a suppression of the classical radiation 
by the quantum effect.  
\end{abstract}


\maketitle

\section{Introduction}
The Unruh effect predicts that an observer in an accelerated 
motion sees the Minkowski vacuum as thermally excited \cite{Unruh}.
Arguments of experimentally detecting the Unruh effect have been 
under debate. 
One such argument is initiated the work by Chen and Tajima 
\cite{ChenTajima,ELI}, who proposed a possible detectable signal 
in the radiation from a charged particle in an accelerated motion, 
which can be realized in an intense laser field. 
However, it has not been clarified whether the radiation originated
from the Unruh effect, which we call the quantum radiation, really 
exists or not \cite{detector,IYZ}. 
The purpose of the present paper is to resolve this problem. 

The authors of Ref.~\cite{Johnson:2005pf} have considered a theoretical model, consisting 
of a particle and a quantum field, which are coupled to each other. 
They have developed a theoretical framework to investigate the properties of 
the random motion of a particle around a classical uniformly accelerated 
motion, which are caused by the coupling to quantum vacuum fluctuations, 
as well as the radiation from the particle in the random motion. 
They have derived a Langevin-like equation for a particle due to the 
coupling to the quantum field, which induces random forces from the 
quantum field fluctuations, and including the radiation reaction force. 
It is found that the random motion of a particle in the transverse direction, 
perpendicular to the direction of the acceleration, satisfies the 
energy equipartition relation (see Refs.~\cite{IYZ,OYZ}, cf. Ref.~\cite{Caceres}). 

In the present paper, we investigate the quantum radiation from the 
random transverse motions of a particle caused by the coupling to
the vacuum fluctuations. The method is an application of the framework
developed in Ref.~\cite{IYZ}, and we evaluate the expectation value 
of the energy momentum tensor of the field coupled to the particle. 
We demonstrate that the energy flux of the quantum radiation is negative 
of the order smaller than that of the classical radiation by the factor $a/m$, where 
$a$ is the acceleration constant and $m$ is the mass of a particle. 
Throughout this paper we adopt the natural unit and 
follow the metric convention $(+,-,-,-)$.

\section{Review of Basic Formulas}

We briefly review the basic formulas and the previous results. 
Following the previous works \cite{IYZ,OYZ}, we consider the system consisting of a 
particle and a scalar field coupled to each other, the action of which is given by
\begin{equation}
 S=S_{\rm P}(z)+S_{\phi}(\phi)+S_{\rm int}(z,\phi),
\label{action2}
\end{equation}
where $S_{\rm P}(z)$ and $S_{\phi}(\phi)$ are the action for the free particle and field, 
\begin{eqnarray}
 &&S_{\rm P}(z)=-m\int d\tau \sqrt{\eta_{\mu\nu} {\dot z}^\mu \dot z^\nu},
~~~~~~~
S_{\phi}(\phi) =\int {d^4x}  \frac{1}{2} \partial^\mu\phi  \partial_\mu\phi,
\label{1-2}
\end{eqnarray}
and $S_{\rm int}(z,\phi)$ describes the interaction, 
\begin{eqnarray}
 &&S_{\rm int}(z,\phi) 
  =  e\int d\tau {d^4x} \sqrt{g_{\mu\nu}(x) {\dot z}^\mu \dot z^\nu} \phi(x) \delta^{4}\left(x-z(\tau)\right)
= e\int d\tau \sqrt{\eta_{\mu\nu}{\dot z}^\mu \dot z^\nu} \phi(z(\tau)). 
\label{1-3}
\end{eqnarray}
where $e$ is the charge of the particle. 
Note that $x^\mu=z^\mu(\tau)$ denotes the trajectory of a particle, which 
obeys
\begin{eqnarray}
 m{ {\ddot z}^\mu}=e\left({ {\ddot z}^\mu} \phi
 +{\dot z}^\mu {\dot z}^\alpha{\partial \phi \over \partial x^\alpha}
 -\eta^{\mu\alpha}{\partial \phi \over \partial x^\alpha}\right)\bigg|_{x=z(\tau)}+F^\mu,
\end{eqnarray}
where $F^\mu$ is a force for a uniformly accelerated motion, while
the equation of motion for the scalar field is
\begin{eqnarray}
  &&\partial^\mu \partial_\mu \phi(x) 
   =e \int d\tau \sqrt{\eta_{\mu\nu}{\dot z}^\mu \dot z^\nu} 
 \delta^4 (x-z(\tau)).
\label{fieldeq}
\end{eqnarray}
The field equation has the solution, 
\begin{eqnarray}
\phi(x)=\phi_{\rm h}(x)+\phi_{\rm inh}(x), 
\end{eqnarray}
where 
$\phi_{\rm h}$ and $\phi_{\rm inh}$ are the homogeneous solution and the inhomogeneous 
solution, respectively. The homogeneous solution satisfies $\partial^\mu\partial_\mu\phi_{\rm h}=0$, 
which we regard as the quantized vacuum field, while the inhomogeneous solution is written as 
\begin{eqnarray}
\phi_{\rm inh}(x)&=&\int d^4x' G_R(x,x')e\int d\tau' \sqrt{\eta_{\mu\nu}{\dot z}^\mu \dot z^\nu} 
\delta^4 (x'-z(\tau'))
=e \int^\tau d\tau' G_R(x,z(\tau')),
\end{eqnarray}
where  $G_R(x,y)$ denotes the retarded Green function  satisfying 
$
\left( \partial^\mu \partial_\mu\right) G_R(x) 
   =\delta^4(x).
$
The term of the inhomogeneous solution $\phi_{\rm inh}$ gives rise to a 
radiation reaction force, and we have 
the stochastic equation of motion \cite{LH,GHL,Hobbs},
\begin{eqnarray}
 &&m{\ddot z^\mu}={e^2 \over 12\pi} \left({\dddot z^\mu}+\dot z^\mu
\Bigl({\ddot z}\Bigr)^2\right)
+e
\left({{\ddot z}^\mu} \phi_{\rm h}
 +{\dot z}^\mu {\dot z}^\alpha{\partial\phi_{\rm h} \over \partial x^\alpha}
 -\eta^{\mu\alpha}{\partial \phi_{\rm h} \over \partial x^\alpha}\right)\bigg|_{x=z(\tau)}+F^\mu.
\label{stochasticeq}
\end{eqnarray}
This stochastic equation of a particle is derived in Ref.~\cite{IYZ}.
We consider a particle in an accelerated motion with a uniform acceleration 
$a$ in the absence of the coupling to the quantum field. 
The equation of motion for random motions around the classical motion is
solved by using the following perturbative method. 
\def\barz{{\bar z}}
Assuming that the trajectory of a particle is written as 
\begin{eqnarray}
z^\mu=\bar z^\mu +\delta z^\mu,
\label{pertz}
\end{eqnarray}
where $\bar z^\mu=(a^{-1}\sinh a\tau,a^{-1}\cosh a\tau,0,0)$ describes the
classical trajectory with a uniformly acceleration, and $\delta z^\mu$
does the random motion due to the coupling to the quantum field. 
Since the {\it transverse} motions satisfy the energy equipartition relation, then 
we consider the perturbative equation of motion for the transverse fluctuations 
\cite{IYZ},
\begin{eqnarray}
&&m\ddot {\delta z}^i={e^2\over 12\pi}(\dddot {\delta z}^i-a^2 \dot {\delta z}^i)
 +{e}{\partial \phi_{\rm h} \over \partial x^i}\Bigr|_{x=z(\tau)}. 
 \label{deltaz}
\end{eqnarray}
The thermal property of the random motions, which are obtained as solutions of 
this equation, has been demonstrated in Ref.~\cite{IYZ,OYZ}.

In the present paper, for simplicity, we drop the third-order time derivative term of 
the radiation reaction force.
As discussed in Appendix A, the contribution of this term to the solution of $\delta z^i$ 
is small, which is suppressed by the order of  
${\cal O}\bigl((a/m)^2\bigr)$.
It is also shown that the contribution comes from the dynamics in small scale about the 
classical electron radius, $r_e= e^2/m$, which is much smaller than the Compton length.
Assumption of the point particle is no longer valid in order to describe such small 
scale behaviors, where one needs to use a more sophisticated model on the basis of the wave 
packet~\cite{Zhang:2013ria}.
Hence we ignore such a term in our description of the point particle.
Now we have
\begin{eqnarray}
m \delta \ddot z^i=-{e^2a^2\over 12\pi}\delta \dot z^i 
+e{\partial \phi_{\rm h}\over \partial x^i}\bigg|_{x=z(\tau)}.
\label{eqofm}
\end{eqnarray}
As will be discussed in the next section, the solutions of this equation exhibit 
the thermal property that the transverse motions satisfy the energy equipartition relation. 
Therefore, we expect that the quantum radiation from the random motions of a particle can be 
investigated if it existed.

The solution of (\ref{eqofm}) with the initial condition $\delta\dot z^i=\delta\dot z^i(\tau_0)$ at the initial time $\tau_0$ is
\begin{eqnarray}
\delta \dot z^i(\tau)=\biggl[\delta \dot z^i(\tau_0) e^{-a\sigma(\tau-\tau_0)}+{e\over m}\int_{\tau_0}^\tau d\tau'\partial_i\phi_{\rm h}(z(\tau'))
e^{-a\sigma(\tau-\tau')}\biggr]\theta(\tau-\tau_0),
\label{solvv}
\end{eqnarray}
with the dimensionless parameter defined by
\begin{eqnarray}
\sigma={e^2a\over 12\pi m}.
\end{eqnarray} 
Introducing the Fourier expansion, 
\begin{eqnarray}
\partial_i\phi_{\rm h}(z(\tau))={1\over 2\pi}\int d\omega
\partial_i\varphi(\omega) e^{-i\omega\tau},
\label{fouri1}
\end{eqnarray}
Eq.~(\ref{solvv}) is rewritten as 
\begin{eqnarray}
\delta \dot z^i(\tau)=\delta \dot z^i(\tau_0) e^{-a\sigma(\tau-\tau_0)}+{e\over 2\pi m}
\int d\omega {\partial_i\varphi(\omega)\over a\sigma -i\omega}
\Bigl\{e^{-i\omega\tau}-e^{-a\sigma(\tau-\tau_0)-i\omega\tau_0} \Bigr\}.
\label{fouri2}
\end{eqnarray}

The retarded Green function for the massless scalar field is written as
$G_R(x-y)=\theta(x^0-y^0)\delta_D((x-y)^2)/2\pi$, where $\delta_D(z)$ denotes the 
Dirac delta function. Using this retarded Green function we have 
\begin{eqnarray}
\phi_{\rm inh}(x)=e\int d\tau G_R(x-z(\tau))
={e\over 4\pi \rho(x)}
\end{eqnarray}
with 
\begin{eqnarray}
\rho(x)=\dot z_\mu(\tau^x_-)(x^\mu-z^\mu(\tau_-^x)),
\end{eqnarray}
where $\tau_-^x$ is the solution of $(x-z(\tau_-^x))^2=0$. In the present paper, we
find the solution using the perturbative method. For the particle trajectory, 
assuming $z^\mu=\bar z^\mu+\delta z^\mu$, where $\delta z^\mu$ describes the perturbation around the classical path $\bar z^\mu$,
the function $\rho(x)$ is expanded as 
\begin{eqnarray}
\rho(x)=\rho_0(x)+\delta \rho(x)+\cdots,
\end{eqnarray}
where we defined
\begin{eqnarray}
&&\rho_0(x)={\dot \barz}(\tau_-^x)\cdot(x-\barz(\tau_-^x)),
~~~~
\delta \rho(x)\simeq \delta \dot z(\tau_-^x)\cdot(x-\barz(\tau_-^x)).
\label{deltarho}
\end{eqnarray}
Then, up to the first order of perturbations, the inhomogeneous solution is given by
\begin{eqnarray}
  \phi_{\rm inh}(x)\simeq{e\over 4\pi \rho_0(x)}\biggl(1-{\delta \rho(x)\over \rho_0(x)}\biggr).
\end{eqnarray}

At the first order of perturbations, $\tau_-^x$ is determined by $(x-\bar z(\tau_-^x))^2=0$, 
i.e., 
$ (x^0-\bar z^0(\tau)-i\epsilon)^2
-(x^1-\bar z^1(\tau))^2-x_\perp^2=0$,
with $x^2_\perp=(x^2)^2+(x^3)^2$ under the condition $x^0>z^0(\tau^x_-)$.
The expression of $\tau_-^x$ is  
\begin{eqnarray}
&&
\hspace{-4.cm}
\tau_-^x={{1\over a}\log\left[{a\over 2(x^0-x^1)}\left(-L_x^2+\sqrt{L_x^4+{4\over a^2}[(x^0)^2-(x^1)^2]}
\right)\right]},
\label{deftaum}
\end{eqnarray}
for $x^\mu$ in both the R-region ($t<x^1,~t>-x^1$) and F-region ($t>x^1,~t>-x^1$), 
where $L_x^2$ is defined by $L_x^2=-x^\mu x_\mu+1/a^2$.
The physical meaning of $\tau_-^x$ is understood with Fig. 1.
For the arguments below, we introduce $\tau_+^x$, which 
is the other solution of $(x^0-\bar z^0(\tau)-i\epsilon)^2
-(x^1-\bar z^1(\tau))^2-x_\perp^2=0$, as
\begin{eqnarray}
\tau_+^x=
{{1\over a}\log\left[{a\over 2(x^0-x^1)}\left(\mp L_x^2 \mp \sqrt{L_x^4+{4\over a^2}[(x^0)^2-(x^1)^2]}
\right)\right]} ~~~~&{\rm for } ~x^\mu{\rm ~in~\left( 
\begin{array}{c}
{\rm R} \\
{\rm F}
\end{array}
\right)
~region},
\label{deftaup}
\end{eqnarray}
respectively. The physical meaning of $\tau_+^x$ is represented with Fig.~1.
\begin{figure}[b]
 \begin{minipage}{0.5\hsize}
  \begin{center}
  \vspace{ .8cm}
   \hspace{-14.cm}
   \includegraphics[width=80mm,bb=0 0 640 480]{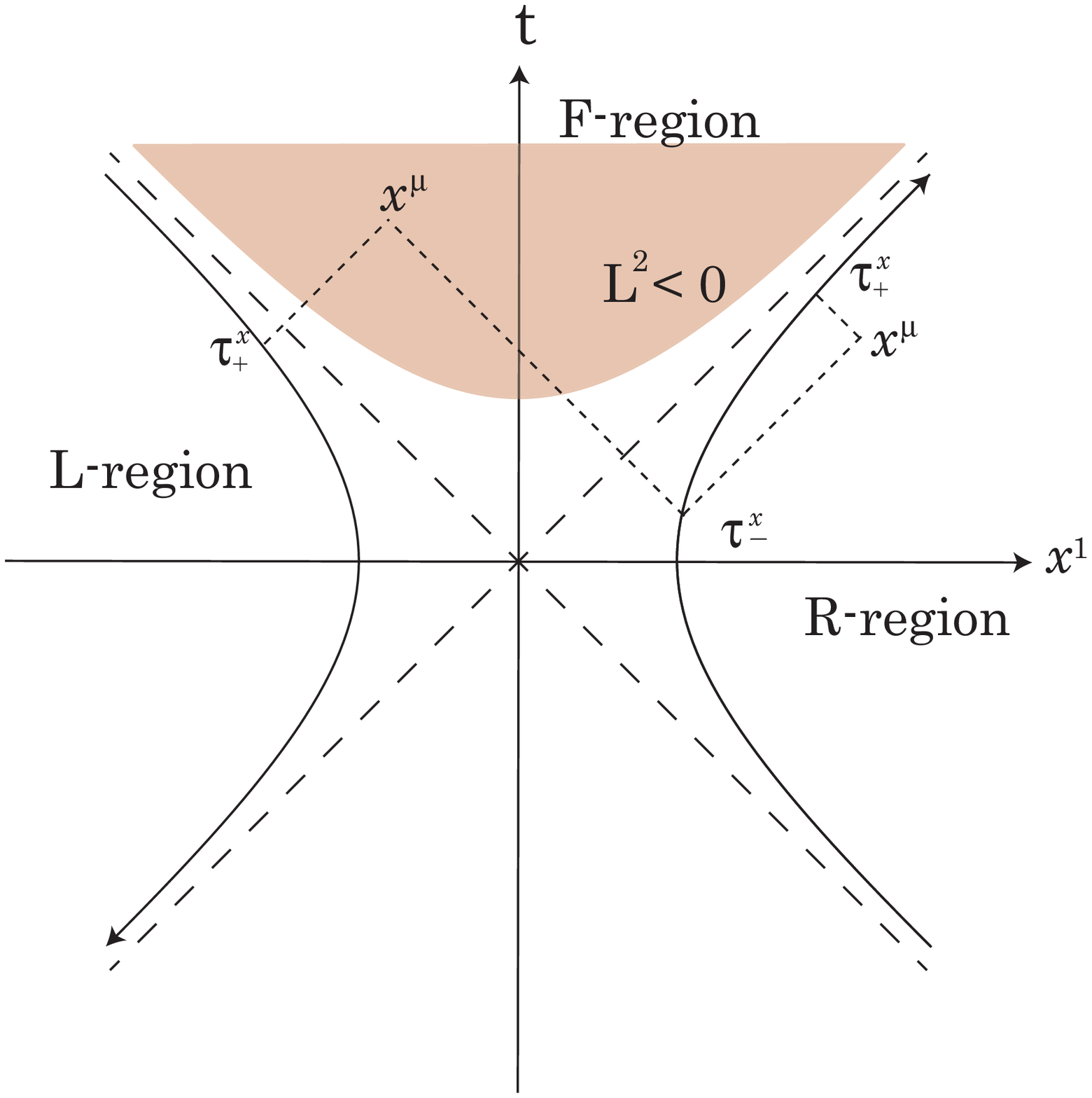}
  \end{center}
  \label{fig:two}
 \end{minipage}
 \caption{A sketch of the configuration of the coordinate.  
The hyperbolic curve in the R-region is the trajectory of 
a uniformly accelerated particle. The hyperbolic curve in 
the L-region is the hypothetical trajectory obtained by an
analytic continuation of the true trajectory. 
When an observer's point $x^\mu$ is in the R-region, $\tau_\pm^x$ is
defined by the proper time of the particle's trajectory intersecting
with the future and past light cone, respectively.
When an observer's point $x^\mu$ is in the F-region, the $\tau_-^x$ is
defined in the same way, but $\tau_+^x$ is the proper time of the 
hypothetical trajectory in the L-region intersecting
with the past light cone. The shaded region satisfies $L_x^2<0$. 
}
\end{figure}

To evaluate the energy momentum tensor,
we first consider two-point function
\begin{eqnarray}
\langle\phi(x)\phi(y)\rangle=\langle\phi_{\rm h}(x)\phi_{\rm h}(y)\rangle
+\langle\phi_{\rm h}(x)\phi_{\rm inh}(y)\rangle
+\langle\phi_{\rm inh}(x)\phi_{\rm h}(y)\rangle
+\langle\phi_{\rm inh}(x)\phi_{\rm inh}(y)\rangle.
\end{eqnarray}
The interference term and the inhomogeneous term are expressed as 
\begin{eqnarray}
&&\langle \phi_{\rm inh} (x) \phi_{\rm h} (y) \rangle + \langle \phi_{\rm h} (x) \phi_{\rm inh} (y) \rangle
= - \frac{e}{4 \pi} \left( \frac{\langle \delta \rho (x) \phi_{\rm h} (y) \rangle}{\rho_0^2(x)} + \frac{\langle \phi_{\rm h} (x) \delta \rho (y) \rangle}{\rho_0^2 (y)} \right),
\label{interference}
\\
&&\langle \phi_{\rm inh} (x) \phi_{\rm inh} (y) \rangle
= \biggl(\frac{e}{4 \pi\rho_0 (x) }\biggr)^2 \left( 1+{\langle\delta\rho(x)\delta\rho(y)\rangle
\over \rho_0(x)\rho_0(y)}\right),
\label{inhomogeneous}
\end{eqnarray}
respectively. We first consider the interference term. 
Note that $\langle \delta \rho (x) \phi_{\rm h} (y) \rangle$ is given by
\begin{eqnarray}
\langle \delta \rho (x) \phi_{\rm h} (y) \rangle &=& \langle \delta \dot{z}^\mu(\tau_{-}^{x})(x_\mu-\bar z_\mu 
(\tau_{-}^{x}))  \phi_{\rm h} (y) \rangle
\nonumber
\\
&=& -x^i\langle\delta \dot{z}^i(\tau_{-}^{x}) \phi_{\rm h} (y) \rangle, 
\end{eqnarray}
for the transverse fluctuations. Here the symbol $\sum_{i=2,3}$ is omitted in the second line 
of this expression.
Assuming $\langle\delta \dot z(\tau_0) \phi_{\rm h}(y)\rangle=0$, 
which means the detector and the scalar field do not 
correlate to each other at the initial time $t=z^0(\tau_0)$, 
we have
\begin{eqnarray}
\langle \delta \rho (x) \phi_{\rm h} (y) \rangle &=&
-{e\over m} x^i \int_{\tau_0}^{\tau_-^x} d\tau e^{-a\sigma(\tau_-^x-\tau)}
\langle\partial_i\phi_{\rm h}(z(\tau))\phi_{\rm h}(y)\rangle\theta(\tau_-^x-\tau_0).
\end{eqnarray}
Using ~(\ref{fouri1}) or (\ref{fouri2}), we have
\begin{eqnarray}
\langle \delta \rho (x) \phi_{\rm h} (y) \rangle &=&
-{e\over m} {x^i\over 2\pi}\int d\omega
\langle\partial_i\varphi(\omega)\phi_{\rm h}(y)\rangle
{e^{-i\omega\tau_-^x}-e^{-a\sigma\tau_-^x+(a\sigma-i\omega)\tau_0}\over a\sigma-i\omega}
\theta(\tau_-^x-\tau_0),
\end{eqnarray}
which leads to
\begin{eqnarray}
&&\hspace{-1cm}\langle \phi_{\rm inh} (x) \phi_{\rm h} (y) \rangle + \langle \phi_{\rm h} (x) \phi_{\rm inh} (y) \rangle
\nonumber\\
&&={e^2\over 4\pi m}\biggl[{x^i\over \rho_0^2(x)}\int{d\omega\over 2\pi} 
\langle\partial_i\varphi(\omega)\phi_{\rm h}(y)\rangle
{e^{-i\omega\tau_-^x}-e^{-a\sigma\tau_-^x+(a\sigma-i\omega)\tau_0}\over a\sigma-i\omega}
\theta(\tau_-^x-\tau_0)
\nonumber\\
&&~~~~~~~~+{y^i\over \rho_0^2(y)}\int{d\omega\over 2\pi} 
\langle\phi_{\rm h}(x)\partial_i\varphi(\omega)\rangle
{e^{-i\omega\tau_-^y}-e^{-a\sigma\tau_-^y+(a\sigma-i\omega)\tau_0}\over a\sigma-i\omega}
\theta(\tau_-^y-\tau_0)\biggr].
\label{pp}
\end{eqnarray}
The right-hand side of (\ref{pp}) can be evaluated as, using the relation found in
Ref.~\cite{IYZ},
\begin{eqnarray}
&&\braket{\partial_i \varphi(\omega) \phi_{\rm h}(y)} = -\frac{iay^i}{4\pi \rho_0^2(y)}\frac{1}{e^{-2\pi \omega/a}-1} \left( \left( \frac{aL_y^2}{2\rho_0(y)} + \frac{i\omega}{a} \right) e^{i\omega \tau_-^y}+ \left( -\frac{aL_y^2}{2\rho_0(y)} + \frac{i\omega}{a} \right) e^{i\omega \tau_{+}^{y}} Z_y(-\omega)\right),
\label{wwxxx}
\\
&&\braket{\phi_{\rm h}(x) \partial_i \varphi(\omega)}=\frac{iax^i}{4\pi \rho_0^2(x)}\frac{1}{e^{2\pi \omega/a}-1} \left( \left( \frac{aL_x^2}{2\rho_0(x)} + \frac{i\omega}{a} \right) e^{i\omega \tau_-^x}+ \left( -\frac{aL_x^2}{2\rho_0(x)} + \frac{i\omega}{a} \right) e^{i\omega \tau_{+}^{x}} Z_x(\omega)\right). 
\label{xxwww}
\end{eqnarray}

The expression of $\langle \phi_{\rm inh} (x) \phi_{\rm inh} (y) \rangle$ is given as
\begin{eqnarray}
&&\hspace{-1cm}\langle \phi_{\rm inh} (x) \phi_{\rm inh} (y) \rangle=
{e^2\over (4\pi)^2}{1\over \rho_0(x)\rho_0(y)}\biggl[1+{x^iy^j \langle 
\delta\dot z^i(\tau_0)\delta\dot z^j(\tau_0)\rangle \over \rho_0(x)\rho_0(y)}e^{-a\sigma(\tau_-^x-\tau_0)}e^{-a\sigma(\tau_-^y-\tau_0)}
\nonumber\\
&&
+
{e^2\over m^2}{x^i y^i\over \rho_0(x)\rho_0(y)}
\int {d\omega \over 2\pi }{a^3\over 12\pi^2}
{e^{-i\omega\tau_-^x}-e^{-a\sigma\tau_-^x+(a\sigma-i\omega)\tau_0}\over a\sigma-i\omega}
{e^{+i\omega\tau_-^y}-e^{-a\sigma\tau_-^y+(a\sigma+i\omega)\tau_0}\over a\sigma+i\omega}
\nonumber\\
&&
\biggr]\theta(\tau_-^x-\tau_0)\theta(\tau_-^y-\tau_0).
\label{pinhpinh1}
\end{eqnarray}
Note that we can omit the second term of the right-hand side of (\ref{pinhpinh1}),
which depends on the initial velocity, by setting $\langle\delta\dot z^i(\tau_0)\delta\dot z^j(\tau_0)\rangle=0$. 

Adding (\ref{pinhpinh1}) to (\ref{pp}), a cancellation occurs (Ref.~\cite{IYZ}, 
cf., Ref.~\cite{IYZ13}), and we have
\begin{eqnarray}
&&\hspace{-1cm}\langle \phi_{\rm inh} (x) \phi_{\rm h} (y) \rangle + \langle \phi_{\rm h} (x) \phi_{\rm inh} (y) \rangle+\langle \phi_{\rm inh} (x) \phi_{\rm inh} (y) \rangle=
{e^2\over (4\pi)^2}{1\over \rho_0(x)\rho_0(y)}+
{-iae^2\over m(4\pi)^2}{x^i\over \rho_0^2(x)}{y^i\over \rho_0^2(x)}\biggl[
\nonumber\\
&&~~~~~
\int{d\omega\over 2\pi} 
\frac{1}{e^{-2\pi \omega/a}-1} \left( \left( \frac{aL_y^2}{2\rho_0(y)} \right) e^{i\omega \tau_-^y}+ \left( -\frac{aL_y^2}{2\rho_0(y)} + \frac{i\omega}{a} \right) e^{i\omega \tau_{+}^{y}} Z_y(-\omega)\right)
\nonumber\\
&&~~~~~\times{e^{-i\omega\tau_-^x}-e^{-a\sigma\tau_-^x+(a\sigma-i\omega)\tau_0}\over a\sigma-i\omega}
\theta(\tau_-^x-\tau_0)
\nonumber\\
&&~~~-\int{d\omega\over 2\pi} 
\frac{1}{e^{2\pi \omega/a}-1} \left( \left( \frac{aL_x^2}{2\rho_0(x)} \right) e^{i\omega \tau_-^x}+ \left( -\frac{aL_x^2}{2\rho_0(x)} + \frac{i\omega}{a} \right) e^{i\omega \tau_{+}^{x}} Z_x(\omega)\right)
\nonumber\\
&&~~~~~\times{e^{-i\omega\tau_-^y}-e^{-a\sigma\tau_-^y+(a\sigma-i\omega)\tau_0}\over a\sigma-i\omega}
\theta(\tau_-^y-\tau_0)\biggr],
\end{eqnarray}
which is equivalent to
\begin{eqnarray}
&&\hspace{-1cm}\langle \phi_{\rm inh} (x) \phi_{\rm h} (y) \rangle + \langle \phi_{\rm h} (x) \phi_{\rm inh} (y) \rangle+\langle \phi_{\rm inh} (x) \phi_{\rm inh} (y) \rangle=
{e^2\over (4\pi)^2}{1\over \rho_0(x)\rho_0(y)}+
{-iae^2\over m(4\pi)^2}{x^i\over \rho_0^2(x)}{y^i\over \rho_0^2(x)}\biggl[
\nonumber\\
&&~~~~~
\int{d\omega\over 2\pi} 
\frac{1}{e^{-2\pi \omega/a}-1} \left( \left( \frac{aL_y^2}{2\rho_0(y)} \right) e^{i\omega \tau_-^y}+ \left( -\frac{aL_y^2}{2\rho_0(y)} + \frac{i\omega}{a} \right) e^{i\omega \tau_{+}^{y}} Z_y(-\omega)\right)
\nonumber\\
&&~~~~~\times{e^{-i\omega\tau_-^x}-e^{-a\sigma\tau_-^x+(a\sigma-i\omega)\tau_0}\over a\sigma-i\omega}
\theta(\tau_-^x-\tau_0)
\nonumber\\
&&~~~-\int{d\omega\over 2\pi} 
\frac{1}{e^{-2\pi \omega/a}-1} \left( \left( \frac{aL_x^2}{2\rho_0(x)} \right) e^{-i\omega \tau_-^x}+ \left( -\frac{aL_x^2}{2\rho_0(x)} - \frac{i\omega}{a} \right) e^{-i\omega \tau_{+}^{x}} Z_x(-\omega)\right)
\nonumber\\
&&~~~~~\times{e^{+i\omega\tau_-^y}-e^{-a\sigma\tau_-^y+(a\sigma+i\omega)\tau_0}\over a\sigma+i\omega}
\theta(\tau_-^y-\tau_0)\biggr].
\end{eqnarray}
Thus, the quantum part of $\langle \phi_{\rm inh}(x) \phi_{\rm inh}(y) \rangle$ cancels out, and the 
classical part of it contributes to the classical radiation, the first term of the rhs 
of the 
above expression. Therefore, the quantum radiation comes from the remaining interference terms
in 
$\langle \phi_{\rm inh}(x) \phi_{\rm h}(y) \rangle+\langle \phi_{\rm h}(x) \phi_{\rm inh}(y) \rangle$. 
Since the quantum interference is the origin of the quantum radiation, it might be difficult 
to understand their properties in an intuitive manner, simply expected as the radiation from a particle. 
Therefore, the appearance of the thermal properties in the quantum radiation is not 
guaranteed although the random motions of the particle exhibit the thermal property. 

\section{Energy equipartition relation}
We demonstrate the energy equipartition relation for the particle's random motions. 
The velocity of the random motion is defined by $v^i(\tau)=\delta\dot z^i(\tau)$. 
From Eq.~(\ref{solvv}), the two-point function of the velocity is given by
\begin{eqnarray}
\langle v^i(\tau)v^j(\tau')\rangle&=&
{e^2\over m^2}\int_{\tau_0}^\tau \int_{\tau_0}^{\tau'} d\tau'' d\tau'''
e^{-a\sigma(\tau-\tau'')}e^{-a\sigma(\tau'-\tau''')}\langle\partial_i\phi_{\rm h}(z(\tau''))
\partial_j\phi_{\rm h}(z(\tau'''))\rangle,
\label{vvvvv}
\end{eqnarray}
where we omitted the term $\langle v^i(\tau_0)v^j(\tau_0)\rangle$.
We follow the argument in Ref.~\cite{IYZ}. Using the expressions,
\begin{eqnarray}
&&\Bigl<\phi_{\rm h}(x)\phi_{\rm h}(x')\Bigr>=-{1\over 4\pi^2}{1\over (t-t'-i\epsilon)^2-|{\bf x-x}'|^2},
\\
&&\Bigl<\partial_i\phi_{\rm h}(z(\tau))\partial_j\phi_{\rm h}(z(\tau'))\Bigr>
={a^4\over 32\pi^2}{\delta_{ij}\over \sinh^4\Bigl({a(\tau-\tau'-i\epsilon)/ 2}\Bigr)},
\end{eqnarray}
where $\epsilon$ is a small positive constant, Eq.~(\ref{vvvvv}) yields
\begin{eqnarray}
&&\Bigl<v^i(\tau)v^j(\tau')\Bigr>=
{e^2\delta_{ij}\over 12\pi^2 m^2}
\int d\omega {1\over a^2\sigma^2+\omega^2}{\omega(\omega^2+a^2)\over 1-e^{-2\pi\omega/a}}  
\bigl(e^{-i\omega\tau}-e^{-a\sigma(\tau-\tau_0)-i\omega\tau_0}\bigr)
\bigl(e^{+i\omega\tau'}-e^{-a\sigma(\tau'-\tau_0)+i\omega\tau_0}\bigr),
\nonumber\\
\end{eqnarray}
where we used (\ref{fouri1}) and $\left<\partial_i\varphi(\omega)\partial_j\varphi(\omega')\right>
=2\pi \delta_D(\omega+\omega')\delta_{ij}(\omega^3+\omega a^2)/(6\pi(1-e^{-2\pi \omega/a}))$.
In the limit of $\tau_0\rightarrow -\infty$, the 
two-point function which is symmetrized with respect to $\tau$ and $\tau'$ reduces to
\begin{eqnarray}
&&\bigl<v^i(\tau)v^j(\tau')\bigr>_S=
{e^2\delta_{ij}\over 24\pi^2 m^2}
\int d\omega \omega{\omega^2+a^2\over a^2\sigma^2+\omega^2} \coth (\pi \omega/a) e^{i\omega(\tau-\tau')}.
\end{eqnarray}
The poles of the integrand in the complex plane of $\omega$ is $\pm a\sigma$, and 
$\pm ian$ with $(n=2,3,4,\cdots)$, and we have
\begin{eqnarray}
\Bigl<v^i(\tau)v^j(\tau')\Bigr>_S={e^2\delta_{ij}\over 24\pi^2 m^2}\left\{
\pi a^2(1-\sigma^2)\cot\pi\sigma e^{-a\sigma|\tau-\tau'|}
-2a^2\sum_{n=2}^\infty{n(n^2-1)\over n^2-\sigma^2}e^{-na|\tau-\tau'|}
\right\}.
\end{eqnarray}
The first term in the rhs of the above equation 
comes from the low energy pole $\omega=\pm ia\sigma$, 
while the  latter terms come from the thermal poles $\omega=\pm ina$.
After summing up the terms from the thermal poles, we have
\begin{eqnarray}
\Bigl<v^i(\tau)v^j(\tau')\Bigr>_S\simeq{\delta_{ij}}{a\over 2\pi m}
-\delta_{ij} {a^2e^2\over 12\pi^2 m^2}\biggl\{{1\over (a|\tau-\tau'|)^2}+\log |a(\tau-\tau')|\biggr\}
.
\end{eqnarray}
The low energy pole leads to the first term in the rhs of this expression, 
which represents the energy equipartition relation with the 
Unruh temperature $T_U=a/2\pi$, while the 
thermal poles give the second term. The thermal poles only give the term of higher 
order of the power of $\sigma^2={\cal O}(a^2/m^2)$ but include the divergence in the limit of
the coincidence limit. One may understand that this divergence comes from the 
short-distance motion of the particle, originated from our formulation based on the point 
particle \cite{OYZ}. The divergence coming from the short-distance motion of the particle 
will be 
removed by taking a finite size effect of the particle into account. Therefore, this 
suggests that $|\tau-\tau'|$ cannot be taken to be zero, and it is natural to introduce a finite value cutoff.

\section{Energy Momentum Tensor}
We consider the energy momentum tensor in the limit of $\tau_0\rightarrow -\infty$, 
which is derived from the two-point function,
\begin{eqnarray}
&&\hspace{-1cm}\langle \phi_{\rm inh} (x) \phi_{\rm h} (y) \rangle + \langle \phi_{\rm h} (x) \phi_{\rm inh} (y) 
\rangle+\langle \phi_{\rm inh} (x) \phi_{\rm inh} (y) \rangle={e^2\over (4\pi)^2}{1\over \rho_0(x)\rho_0(y)}
+{-iae^2\over m(4\pi)^2}{x^i\over \rho_0^2(x)}{y^i\over \rho_0^2(x)}\biggl[
\nonumber\\
&&
+\int{d\omega\over 2\pi} 
\frac{1}{e^{-2\pi \omega/a}-1} \left( \frac{aL_y^2}{2\rho_0(y)} e^{i\omega (\tau_-^y-\tau_-^x)}
+\left( -\frac{aL_y^2}{2\rho_0(y)} + \frac{i\omega}{a} \right) e^{i\omega (\tau_{+}^{y}-\tau_-^x)} Z_y(-\omega)\right)
{1\over a\sigma-i\omega}
\nonumber\\
&&-\int{d\omega\over 2\pi} 
\frac{1}{e^{-2\pi \omega/a}-1} \left( \frac{aL_x^2}{2\rho_0(x)} e^{-i\omega (\tau_-^x-\tau_-^y)}
+ \left( -\frac{aL_x^2}{2\rho_0(x)} - \frac{i\omega}{a} \right) e^{-i\omega (\tau_{+}^{x}-\tau_-^y)} 
Z_x(-\omega)\right)
{1\over a\sigma+i\omega}\biggr].
\label{pppppp}
\end{eqnarray}
This expression is equivalent to that in Ref.~\cite{IYZ} except the following replacement 
of $h(\omega)$,
\begin{eqnarray}
h(\omega)={1\over -im\omega+e^2(\omega^2+a^2)/12\pi}\rightarrow {1\over -im\omega+e^2a^2/12\pi}.
\end{eqnarray}
By symmetrizing the two-point function with respect to $x$ and $y$, we have
\begin{eqnarray}
&&\hspace{-1cm}[\langle \phi_{\rm inh} (x) \phi_{\rm h} (y) \rangle + \langle \phi_{\rm h} (x) \phi_{\rm inh} (y) \rangle+\langle \phi_{\rm inh} (x) \phi_{\rm inh} (y) \rangle]_S={e^2\over (4\pi)^2}{1\over \rho_0(x)\rho_0(y)}
\nonumber\\
&&~~~~~~~~~
+{-iae^2\over 2m(4\pi)^2}{x^i\over \rho_0^2(x)}{y^i\over \rho_0^2(x)}\biggl[
\frac{aL_x^2}{2\rho_0(x)}(I_3(x,y)-I_1(x,y))+{i\over a} I_2(x,y)\biggr]
+(x\leftrightarrow y),
\end{eqnarray}
where we defined
\begin{eqnarray}
&&I_1(x,y)=\int_{-\infty}^{+\infty}{d\omega\over 2\pi}{e^{-i\omega(\tau_-^y-\tau_+^x)}\over a\sigma-i\omega}\biggl\{\biggl(
{e^{\pi \omega/a}\over 1-e^{2\pi(\omega-i\epsilon)/a}}-{e^{-\pi \omega /a}\over 1-e^{-2\pi(\omega+i\epsilon)/a}}
\biggr)\theta(u_x)
\nonumber\\
&&\hspace{4.9cm}+
\biggl({1\over 1-e^{2\pi(\omega-i\epsilon)/a}}-{1\over 1-e^{-2\pi(\omega+i\epsilon)/a}}\biggr)\theta(-u_x)\biggr\},
\\
&&I_2(x,y)=\int_{-\infty}^{+\infty}{d\omega\over 2\pi}{\omega e^{-i\omega(\tau_-^y-\tau_+^x)}\over a\sigma-i\omega}\biggl\{\biggl(
{e^{\pi \omega/a}\over 1-e^{2\pi(\omega-i\epsilon)/a}}-{e^{-\pi \omega /a}\over 1-e^{-2\pi(\omega+i\epsilon)/a}}
\biggr)\theta(u_x)
\nonumber\\
&&\hspace{4.9cm}+
\biggl({1\over 1-e^{2\pi(\omega-i\epsilon)/a}}-{1\over 1-e^{-2\pi(\omega+i\epsilon)/a}}\biggr)\theta(-u_x)\biggr\}, 
\\
&&I_3(x,y)=\int_{-\infty}^{+\infty}{d\omega\over 2\pi}{e^{-i\omega(\tau_-^x-\tau_-^y)}\over a\sigma+i\omega}\biggl(
{1\over 1-e^{-2\pi(\omega+i\epsilon)/a}}+{1\over e^{2\pi(\omega-i\epsilon)/a}-1}
\biggr).
\end{eqnarray}
When performing the above integrals, we introduced the regulator $i\epsilon$ for the 
pole $\omega=0$, but the results do not depend on the sign of the regulator 
(see Appendix B). 
We find the expression by expanding it with respect to $\sigma=e^2a/12\pi m\ll1$. 
\subsection{F-region $u=x^0-x^1>0$}
In the F-region, using the perturbative expansion with respect to 
$\sigma=e^2a/12\pi m$, we have the expressions,
\begin{eqnarray}
&&I_1(x,y)=
-{i\over 2\pi\sigma}+{i\over \pi}\log \Bigl(1+e^{-a|\tau_-^y-\tau_+^x|}\Bigr)
+{{i}\over \pi}a(\tau_-^y-\tau_+^x)\theta(\tau_-^y-\tau_+^x)+{\cal O}(\sigma),
\\
&&I_2(x,y)=-{a\over \pi} {1\over e^{a(\tau_+^x-\tau_-^y)}+1}+{\cal O}(\sigma),
\\
&&I_3(x,y)=
-{i\over 2\pi\sigma}+{i\over \pi}\log \Bigl(1-e^{-a|\tau_-^y-\tau_-^x|}\Bigr)
+{{i}\over \pi}a(\tau_-^y-\tau_-^x)\theta(\tau_-^y-\tau_-^x)+{\cal O}(\sigma),
\end{eqnarray}
and the symmetric two-point function is
\begin{eqnarray}
&&\hspace{-1cm}[\langle \phi_{\rm inh} (x) \phi_{\rm h} (y) \rangle + \langle \phi_{\rm h} (x) \phi_{\rm inh} (y) \rangle+\langle \phi_{\rm inh} (x) \phi_{\rm inh} (y) \rangle]_S={e^2\over (4\pi)^2}{1\over \rho_0(x)\rho_0(y)}
\nonumber\\
&&-
{iae^2\over 2m(4\pi)^2}{x^i\over \rho_0^2(x)}{y^i\over \rho_0^2(x)}\biggl[
\frac{aL_x^2}{2\rho_0(x)}\biggl\{{i\over \pi}\log \Bigl(1-e^{-a|\tau_-^y-\tau_-^x|}\Bigr)
+{i\over \pi}a(\tau_-^y-\tau_-^x)\theta(\tau_-^y-\tau_-^x)
\nonumber\\
&&~~~~~~~~~~
-{i\over \pi}\log \Bigl(1+e^{-a|\tau_-^y-\tau_+^x|}\Bigr)
-{i\over \pi}a(\tau_-^y-\tau_+^x)\theta(\tau_-^y-\tau_+^x)\biggr\}
-{i\over \pi}{1\over e^{a(\tau_+^x-\tau_-^y)}+1}\biggr]
\nonumber\\
&&
+(x\leftrightarrow y). 
\end{eqnarray}
The energy flux can be computed as follows. At the leading order of $1/r^2$ and $\sigma$, we have 
\begin{eqnarray}
T_{0i}(x)&=&\lim_{y\rightarrow x}{\partial \over \partial x^0}{\partial \over \partial y^i}
[\langle \phi_{\rm inh} (x) \phi_{\rm h} (y) \rangle + \langle \phi_{\rm h} (x) \phi_{\rm inh} (y) \rangle+\langle \phi_{\rm inh} (x) \phi_{\rm inh} (y) \rangle]_S\nonumber
\\
&=&T_{0i}^{\rm C}+T_{0i}^{\rm Q},
\end{eqnarray}
where $T_{0i}^{\rm C}$ and $T_{0i}^{\rm Q}$ are the classical part and the quantum part, respectively, defined as
\begin{eqnarray}
&&T^{\rm C}_{0i}={e^2\over (4\pi)^2}{x_0 x_i\over \rho^4_0(x)}P^2,
\\
&&T^{\rm Q}_{0i}={2a^3e^2\over m (4\pi)^3}{{\bf x}_\perp^2 x_0 x_i\over \rho_0^6(x)}\biggl[
-4P(3P^2-1)\biggl\{\log a \varepsilon -\log \Bigl(1+e^{-a|\tau_--\tau_+|}\Bigr)
-a(\tau_--\tau_+)\theta(\tau_--\tau_+)
\biggr\}
\nonumber\\
&&~~~~~
+(3P^2-1)\Bigl\{-{2\over e^{a|\tau_--\tau_+|}+1}(\theta(\tau_+-\tau_-)-\theta(\tau_--\tau_+))
+1-2\theta(\tau_--\tau_+)\Big\}
\nonumber\\
&&~~~~~
+2P^2\Bigl\{-{2\over e^{a|\tau_--\tau_+|}+1}(\theta(\tau_+-\tau_-)-\theta(\tau_--\tau_+))
-1-2\theta(\tau_--\tau_+)\Big\}
\nonumber\\
&&~~~~~
-P\Bigl\{{2\over (a\epsilon)^2}-2{e^{a|\tau_--\tau_+|}\over (e^{a|\tau_--\tau_+|}+1)^2}\Big\}
\nonumber\\
&&~~~~~
-8P^2{1\over e^{a(\tau_+-\tau_-)}+1}+8P{e^{a(\tau_+-\tau_-)}\over (e^{a(\tau_+-\tau_-)}+1)^2}
-2{e^{a(\tau_+-\tau_-)}(e^{a(\tau_+-\tau_-)}-1)\over (e^{a(\tau_+-\tau_-)}+1)^3}\biggr].
\label{000}
\end{eqnarray}
Eq.(\ref{000}) is equivalent to 
\begin{eqnarray}
&&T^{\rm Q}_{0i}={2a^3e^2\over m (4\pi)^3}{{\bf x}_\perp^2 x_0 x_i\over \rho_0^6(x)}\biggl[
-4P(3P^2-1)\biggl\{\log a \varepsilon -\log \Bigl(1+e^{-a|\tau_--\tau_+|}\Bigr)
-a(\tau_--\tau_+)\theta(\tau_--\tau_+)
\biggr\}
\nonumber\\
&&~~~
-{2(9P^2-1)\over e^{a(\tau_+-\tau_-)}+1}+(P^2-1)
-P\biggl\{{2\over (a\varepsilon)^2}-{5\over 2}{1\over \cosh^2 (a(\tau_+-\tau_-)/2)}\biggr\}
-{1\over 2}{\tanh(a(\tau_+-\tau_-)/2)\over \cosh^2(a(\tau_+-\tau_-)/2)}
\biggr],
\label{001}
\end{eqnarray}
with 
\begin{eqnarray}
&&P={-aL_x^2\over 2\rho_0(x)},
\end{eqnarray}
which may be explicitly written with
\begin{eqnarray}
&&{L_x^2}=-(t^2-r^2)+{1\over a^2},
\\
&&\rho_0(x)=\sqrt{\biggl({a\over2}(t^2-r^2)-{1\over 2a}\biggr)^2+t^2-r^2\cos^2\theta}~,
\\
&&a(\tau_+-\tau_-)=\log\Biggl[
{+L_x^2+\sqrt{L_x^4+{4\over a^2}(t^2-r^2\cos^2\theta)}
\over
-L_x^2+\sqrt{L^4_x+{4\over a^2}(t^2-r^2\cos^2\theta)}}
\Biggr].
\end{eqnarray}
Here we defined $\varepsilon=|\tau_-^x-\tau_-^y|$, which diverges in the coincidence limit
of the two-points $x$ and $y$. {We may understand that this divergence comes from the 
short-distance motion of a particle, originated from our formulation based on the point 
particle, as is discussed in the velocity two point function. The divergence coming from the short-distance 
motion of the particle could be removed by taking a finite size effect of the particle into account.

\subsection{R-region $u=x^0-x^1<0$}
In the R-region,  $I_1(x,y)$ and $I_2(x,y)$ are estimated as
\begin{eqnarray}
&&I_1(x,y)=
-{i\over 2\pi\sigma}+{i\over \pi}\log \Bigl(1-e^{-a|\tau_-^y-\tau_+^x|}\Bigr)
+{\cal O}(\sigma),
\\
&&I_2(x,y)={a\over \pi} {1\over e^{a(\tau_+^x-\tau_-^y)}-1}+{\cal O}(\sigma), 
\end{eqnarray}
respectively, while $I_3(x,y)$ is the same as that in the F-region, then we have
\begin{eqnarray}
&&\hspace{-1cm}[\langle \phi_{\rm inh} (x) \phi_{\rm h} (y) \rangle + \langle \phi_{\rm h} (x) \phi_{\rm inh} (y) \rangle+\langle \phi_{\rm inh} (x) \phi_{\rm inh} (y) \rangle]_S={e^2\over (4\pi)^2}{1\over \rho_0(x)\rho_0(y)}
\nonumber\\
&&
+{-iae^2\over 2m(4\pi)^2}{x^i\over \rho_0^2(x)}{y^i\over \rho_0^2(x)}\biggl[
\frac{aL_x^2}{2\rho_0(x)}\biggl\{{i\over \pi}\log \Bigl(1-e^{-a|\tau_-^y-\tau_-^x|}\Bigr)
+{i\over \pi}a(\tau_-^y-\tau_-^x)\theta(\tau_-^y-\tau_-^x)
\nonumber\\
&&~~~~~~~~~~~~~~~~~~~~~~~~~~
-{i\over \pi}\log \Bigl(1-e^{-a|\tau_+^x-\tau_-^y|}\Bigr)\biggr\}
+{i\over \pi}{1\over e^{a(\tau_+^x-\tau_-^y)}-1}\biggr]
+(x\leftrightarrow y). 
\end{eqnarray}
We have the energy momentum tensor component 
\begin{eqnarray}
&&T^{\rm Q}_{0i}={2a^3e^2\over m (4\pi)^3}{{\bf x}_\perp^2 x_0 x_i\over \rho_0^6(x)}\biggl[
-4P(3P^2-1)\biggl\{\log a \varepsilon -\log \Bigl(1-e^{-a|\tau_--\tau_+|}\Bigr)\biggr\}
\nonumber\\
&&~~~~~~~~~
+(3P^2-1)\biggl\{{2\over e^{a|\tau_--\tau_+|}-1}+1\bigg\}
+2P^2\biggl\{{2\over e^{a|\tau_--\tau_+|}-1}-1\bigg\}
\nonumber\\
&&~~~~~~~~~
-P\biggl\{{2\over (a\varepsilon)^2}+2{e^{a|\tau_--\tau_+|}\over (e^{a|\tau_--\tau_+|}-1)^2}\bigg\}
\nonumber\\
&&~~~~~~~~~
+8P^2{1\over e^{a(\tau_+-\tau_-)}-1}-8P{e^{a(\tau_+-\tau_-)}\over (e^{a(\tau_+-\tau_-)}-1)^2}
+2{e^{a(\tau_+-\tau_-)}(e^{a(\tau_+-\tau_-)}+1)\over (e^{a(\tau_+-\tau_-)}-1)^3}\biggr],
\end{eqnarray}
which is rewritten as
\begin{eqnarray}
&&T^{\rm Q}_{0i}={2a^3e^2\over m (4\pi)^3}{{\bf x}_\perp^2 x_0 x_i\over \rho_0^6(x)}\biggl[
-4P(3P^2-1)\biggl\{\log a \varepsilon -\log \Bigl(1-e^{-a|\tau_--\tau_+|}\Bigr)\biggr\}
\nonumber\\
&&~~~~~~~~~
+{(18P^2-2)\over e^{a(\tau_+-\tau_-)}-1}+P^2-1
-P\biggl\{{2\over (a\varepsilon)^2}+{5\over 2}{1\over \sinh^2(a(\tau_+-\tau_-)/2)}\bigg\}
+{1\over 2}{\coth(a(\tau_+-\tau_-)/2)\over \sinh^2(a(\tau_+-\tau_-)/2)}
\biggr].
\end{eqnarray}
One finds the divergence in the limit that $\varepsilon=|\tau_-^x-\tau_-^y|$ goes to zero,
similar to the case of the F-region. We also find the other divergence in the limit 
that $\tau_+-\tau_-$ goes to zero, which is the limit that the observer approaches 
the particle's classical trajectory.

\section{Discussion}
Note that $T_{0i}^{\rm Q}$ is smaller than the classical part $T_{0i}^{\rm C}$
by the order of $a/m$. $T_{0i}^{\rm Q}$ includes the divergent terms in the coincidence limit $\varepsilon\rightarrow0$ ,
 which needs to be regularized. 
The divergent terms appear due to our theoretical framework based on the point particle. 
It is demonstrated in Sec. III that the two-point function of the velocity includes 
divergent terms in the coincidence limit, 
which reflects the short-distance dynamics of the point particle.
The divergent terms in $T_{0i}^{\rm Q}$ have the same origin, which should be removed by 
taking a finite size effect of the particle into account \cite{OYZ}. 
Furthermore, one can read that the divergent terms are odd functions of $P$.  
This means that the divergent terms contribute to the energy flux as odd functions of 
$t-r$ at a large distance (see below for details), which vanish if one integrates 
them over the time. In the present paper, we simply omit the divergent terms.  

The energy flux in the laboratory frame is related to the energy 
momentum tensor by $f=-T_{0i}n^i$ with $n^i=x^i/r$. 
Here we consider the energy flux in the F-region. 
The energy flux for the classical part and the quantum part are given by
\begin{eqnarray}
&&  f^{\rm C}={1\over r^2}{a^2e^2\over (4\pi)^2}{G(q)\over \sin^4\theta}\theta(t-x^1),
\label{fdeffdefg}
\\
&&  f^{\rm Q}={1\over r^2}{2a^3e^2\over (4\pi)^3m}{F(q)\over \sin^4\theta}\theta(t-x^1),
\label{fdeffdef}
\end{eqnarray}
respectively, where $G(q)$ and $F(q)$ are defined as
\begin{eqnarray}
&& G(q)={{q^2\over (1+q^2)^3}}, 
\\
&& F(q)={1\over (1+q^2)^3}\biggl[-{4q(2q^2-1)\over \sqrt{1+q^2}^3}
\biggl\{-\log \Bigl(1+e^{-a|\tau_--\tau_+|}\Bigr)-a(\tau_--\tau_+)\theta(\tau_--\tau_+)
\biggr\}
\nonumber\\
&&~~~~
{-{2(8q^2-1)\over (1+q^2) (e^{a(\tau_+-\tau_-)}+1)}}\color{black}{-{1\over 1+q^2}
+{q\over \sqrt{1+q^2}}{5\over 2}{1\over \cosh^2 (a(\tau_+-\tau_-)/2)}
-{1\over 2}{\tanh(a(\tau_+-\tau_-)/2)\over \cosh^2(a(\tau_+-\tau_-)/2)}
\biggr]}
\end{eqnarray}
with
\begin{eqnarray}
q(t,r,\theta)={a\over \sin\theta}\Bigl({t-r}-{1\over 2a^2r}\Bigr) \sim {a\over \sin\theta} \left( t-r\right),
\label{Qdef}
\end{eqnarray}
where we assume that the energy flux is observed far from the particle, i.e.,  
$r \gg z^1(\tau_-^x) > 1/a$. 
$G$ and $F$ in Eqs.~(\ref{fdeffdefg}) and (\ref{fdeffdef}) as functions of $q$
determine the energy flux of the classical part and the quantum part, respectively. 
From Eq.~(\ref{Qdef}), we may regard $q$ as the time variable, $t-r$, 
scaled by $\theta$ and $a$, 
which is valid when the observer is located at a large distance from the particle. 
In the derivation of the above expressions, we assumed $t\sim r$ with taking the limit of
$r\rightarrow \infty$, and used following relations
\begin{eqnarray}
&&{-aL_x^2 \over2}={a\over 2}(t^2-r^2)-{1\over 2a} \simeq r\sin\theta q,
\label{approaL2}
\\
&&\rho_0(x)=\sqrt{\left({a\over 2}(t^2-r^2)-{1\over 2a}\right)^2+t^2-r^2\cos^2\theta}
\simeq r\sin\theta\sqrt{1+q^2},
\label{rho00}
\\
&&P= \frac{-aL_x^2}{2 \rho_0(x)} \simeq{q\over \sqrt{1+q^2}}.
\end{eqnarray}
In the F-region, we may use
\begin{eqnarray}
a(\tau_+-\tau_-)=\log\left[{-q+\sqrt{1+q^2}\over +q+\sqrt{1+q^2}}\right].
\label{ATT}
\end{eqnarray}

\begin{figure}[t]
\begin{center}
 \hspace{-1cm}
  \includegraphics[width=90mm]{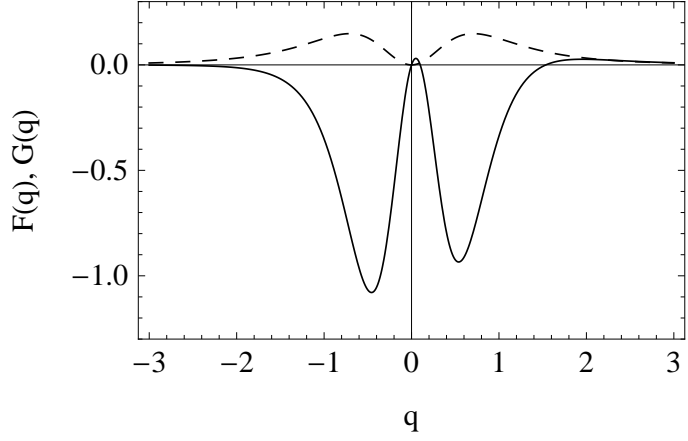}
\caption{Functions $F(q)$ (solid curve) and $G(q)$ (dashed curve).
\label{figure:F}}
  \end{center}
\end{figure}

Figure \ref{figure:F} plots the functions $G(q)$ and $F(q)$. 
The energy flux has the duration of time of the order 
$\Delta t\sim a^{-1}\sin\theta$ for an observer at a large distance.
Also, $F(q)$ is negative in most parts of the region, which means that the energy flux of the quantum part is negative.
To investigate the angular distribution of the energy flux,
we write $q$ as a function of $\tau_-^x$ and $\theta$
\begin{eqnarray}
q(\tau_-^x,\theta) = \sinh{\left[ a \tau_-^x - \text{arctanh}{\left( \cos{\theta} \right)} \right]}.
\end{eqnarray}
Figure \ref{fig:sixx} plots the angular distribution of the classical 
energy flux (left panels) and the quantum energy flux (right panels)
with fixing $\tau_-^x$. 
Explicitly, each left panel in Fig.~\ref{fig:sixx} is the polar plot of 
$\sin^{-4}{\theta}G(q(\tau_-^x, \theta))$, while each right panel is 
$\sin^{-4}{\theta}F(q(\tau_-^x, \theta))$, for $a \tau_-^x=-0.3, ~0, ~0.3$ 
from top to bottom, respectively. 
By this polar plot, each panel represents the energy flux emitted in the direction 
of $\theta$ from the particle at the proper time $\tau=\tau_-^x$.

The classical energy flux in the left panels of Fig.~\ref{fig:sixx}, 
which corresponds to the Larmor radiation in the case for a charged 
particle and the electromagnetic field, has the radiation power in the direction of 
acceleration. 
This is because our model is based on the scalar field 
and the scalar coupling between the particle and the field. 
The longitudinal waves contribute to the classical energy flux in our results unlike the case of 
the electromagnetic field. For a more realistic theoretical prediction, 
we need to consider the model based on the massless vector field 
and the relevant coupling, which will be discussed in a separate paper.
Here one can see the nature of the boostlike behavior for both the classical radiation
and the quantum radiation. 

In the quantum energy flux in the right panels of Fig.~\ref{fig:sixx}, 
the red curve is the negative value.  
But it does not mean that one should observe a negative energy flux from an accelerated 
particle. Only the sum of the classical and the quantum flux is observed. 
The total energy flux is positive as long as $a/m\ll1$.
The results could be understood as a suppression of the classical radiation 
by the quantum effect. The results are consistent with the previous studies
on the quantum correction to the Larmor radiation \cite{HW,NSY,YN,KNY,NY},
though our approach in the present paper is quite different from those previous works. 

\begin{figure}[h]
  \begin{center}
    \vspace{0cm}
   \includegraphics[width=150mm]{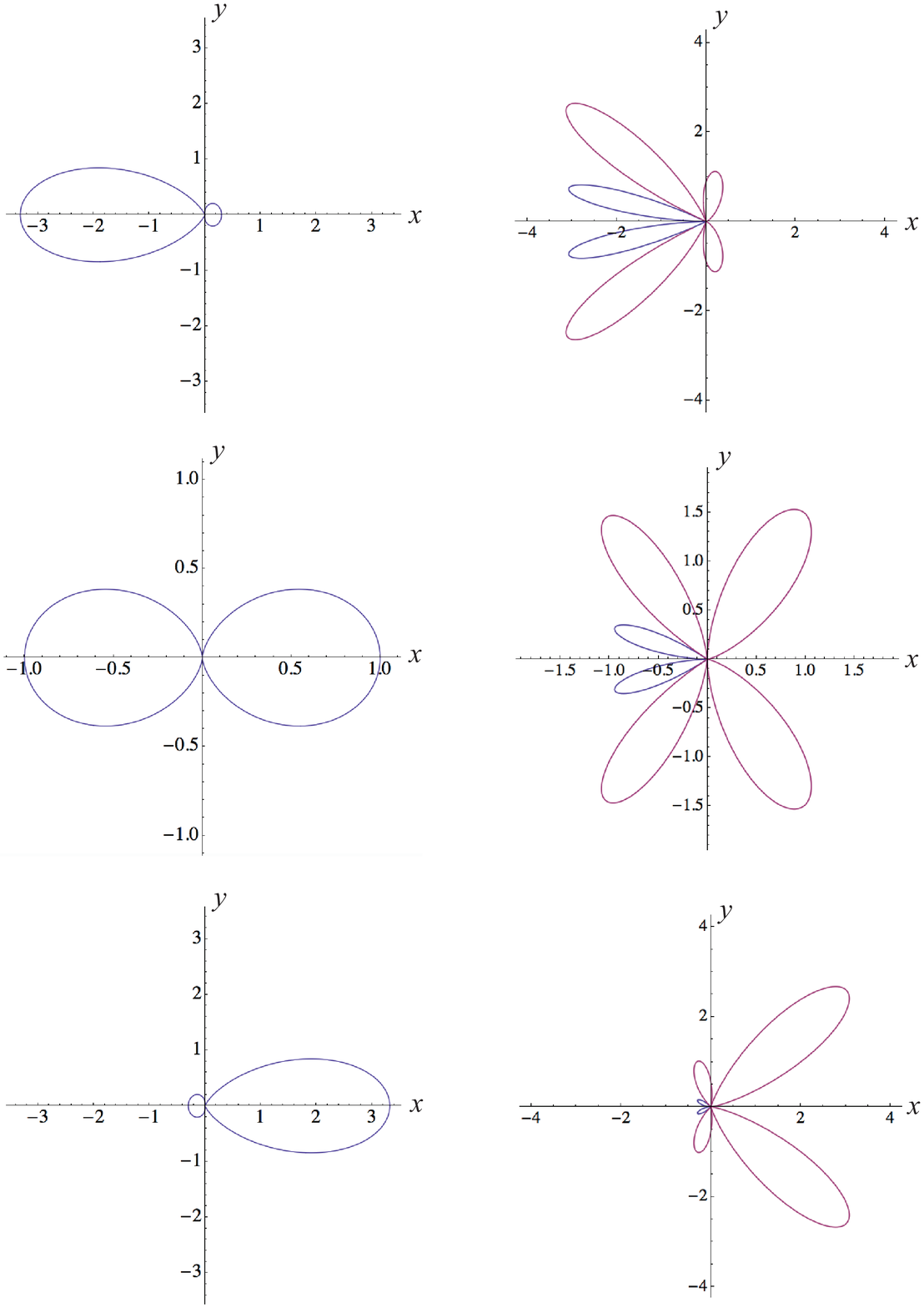}
  \end{center}
   \vspace{0cm}
 \caption{
Angular distribution of the classical radiation $\sin^{-4}{\theta} G(\tau_-^x ,\theta)$ (left panel) 
and the quantum radiation $\sin^{-4}{\theta} F(\tau_-^x ,\theta)$ (right panel) at
$a \tau_-^x = -0.3,~ 0,~0.3$, from top to bottom, respectively. 
The red curve corresponds to a negative value, while the blue curve corresponds to a 
positive value.
\label{fig:sixx}}
\end{figure}

\section{Summary and conclusions}

We have investigated the problem of whether the quantum radiation from a uniformly 
accelerated particle exists or not. 
We adopted the theoretical model consisting of an accelerated particle and a scalar field, 
which was originally developed in Refs.~\cite{IYZ,Johnson:2005pf}. 
We have scrutinized the theoretical features of the energy flux of the field coupled to the 
random thermal motions of an accelerated particle, where we focused on transverse motions 
in the direction perpendicular to the acceleration of the particle, which are demonstrated 
to exhibit the energy equipartition relation. 
Within our model, the energy flux of the radiation is obtained as the sum of the
classical part and the quantum part. The classical part corresponds to the Larmor radiation in the
case of a charged particle and electromagnetic field, but our classical part is not exactly the same
as those of the classical Larmor radiation because of the difference of the model.
However, the quantum part can be considered as the quantum radiation coming from 
the random thermal motions around a uniformly accelerated motion. 
The energy flux of the quantum part is smaller than the
classical part by the order of $a/m$, which shows a unique signature in the angular distribution.
However, the sign of the energy flux of the quantum part is almost negative. The results can be 
understood as a suppression of the total radiation flux by the quantum effect.
This conclusion is consistent with the previous works \cite{HW,NSY,YN,KNY,NY},
which demonstrated that the quantum correction to the Larmor radiation suppresses the total radiation. 
It is quite interesting whether our prediction is the same for the system of the electromagnetic field
and an accelerated charged particle, which will be discussed in a future work. 
The cancellation of the quantum part of $\langle \phi_{\rm inh}(x) \phi_{\rm inh}(y) \rangle$ 
implies that the quantum radiation comes from the quantum interference terms in 
$\langle \phi_{\rm inh}(x) \phi_{\rm h}(y) \rangle+\langle \phi_{\rm h}(x) \phi_{\rm inh}(y) \rangle$. 
This makes it difficult to understand the properties of the quantum radiation intuitively. 
Further investigations are necessary to answer the question whether the quantum 
radiation possesses the thermal properties or not.

\section*{Acknowledgments} 
We would like to thank P. Chen, J. Yokoyama, T. Suyama, Y. Nambu, and K. Fukushima 
for useful discussions.
This work was supported by a financial support program by Hiroshima University 
and by a research program of Advanced Leading Graduate Course for Photon Science 
at the University of Tokyo.
The research by K.Y. is supported in part by a Grant-in-Aid for 
Scientific Research of Japan Ministry of Education, Culture, Sports, 
Science and Technology (No. 15H05895). 


\begin{appendix}

\section{Solution of $\delta \dot{z}^i$}
The solution of Eq.~(\ref{deltaz}) is given by
\begin{eqnarray}
\delta \dot z^i(\tau) &=& 
\biggl[
(\delta \dot z^i(\tau_0) - A ) e^{-\Omega_- (\tau-\tau_0)}
+ A e^{\Omega_+ (\tau-\tau_0)}
\nonumber \\
& & + {12\pi \over e(\Omega_+ + \Omega_-)}\int_{\tau_0}^\infty d\tau'\partial_i\phi_{\rm h}(z(\tau'))
( e^{-\Omega_-(\tau-\tau')}\theta(\tau-\tau') + e^{\Omega_+ (\tau-\tau')} \theta(\tau'-\tau) )
\biggr]\theta(\tau-\tau_0),
\end{eqnarray}
where $A$ is a constant, and $\Omega_\pm$ are defined by
\begin{eqnarray}
\Omega_\pm = \frac{a}{2\sigma} (\sqrt{1+4\sigma^2} \pm 1).
\end{eqnarray}
The third derivative term in the equation, $\delta \dddot z^i$, 
generates the terms with $\Omega_-$ and the additional terms with $\Omega_+$. 
Since $\Omega_-$ is approximated as $\Omega_-=a\sigma(1+{\cal O}(\sigma^2))$, 
the third derivative term contributes to the small correction shifting  
$a\sigma$ to $\Omega_-$.  
Requiring the solution to be regular, one needs to set $A=0$.
The other term inside the integral presents a preacceleration 
feature, reflecting the finite size effects of a particle, 
which is only important on very small scales of the electron radius of 
the order $r_e=e^2/m$.
By expanding $\partial_i \phi_{\rm h}(z(\tau'))$ around $\tau$ as
\begin{eqnarray}
\partial_i \phi_{\rm h}(z(\tau')) = \sum_{k=0}^\infty (\tau'-\tau)^k \frac{d^k}{d\tau^k} \partial_i \phi_{\rm h}(z(\tau)),
\end{eqnarray}
and performing the integration
\begin{eqnarray}
{12\pi \over e(\Omega_+ + \Omega_-)}
\int_{\tau}^\infty d\tau' \ \partial_i\phi_{\rm h}(z(\tau'))
 e^{\Omega_+ (\tau-\tau')} 
\ \sim \ 
{12\pi \over e(\Omega_+ + \Omega_-)} \sum_{k=0}^\infty k! \ \Omega_+^{-k-1} \frac{d^k}{d\tau^k} \partial_i\phi_{\rm h}(z(\tau)) = {12\pi \over e \Omega_+^2 } (\partial_i \phi_{\rm h}(z(\tau)) + {\cal O}(\sigma^2) ) , \nonumber
\end{eqnarray}
one finds the preacceleration term is of the order ${\cal O}(\sigma^2)$.

\section{Integral formula}
In this Appendix, we demonstrate a proper prescription for the following integration:
\begin{eqnarray}
B=\int_{-\infty}^{+\infty} {d\omega\over 2\pi}{e^{ia\omega}-e^{ib\omega}\over \omega}.
\end{eqnarray}
By introducing the regulator $i\epsilon$ with $\epsilon>0$ in the denominator,
\begin{eqnarray}
B=\int_{-\infty}^{+\infty} {d\omega\over 2\pi}{e^{ia\omega}-e^{ib\omega}\over \omega+i\epsilon},
\end{eqnarray}
we have the following expression
\begin{eqnarray}
B=-i\theta(-a)+i\theta(-b).
\end{eqnarray}
We may introduce the regulator in the opposite way
\begin{eqnarray}
B=\int_{-\infty}^{+\infty} {d\omega\over 2\pi}{e^{ia\omega}-e^{ib\omega}\over \omega-i\epsilon},
\end{eqnarray}
which yields
\begin{eqnarray}
B=i\theta(a)-i\theta(b).
\end{eqnarray}
The above results are equivalent to each other because of the relations
$\theta(a)=1-\theta(-a)$ and $\theta(b)=1-\theta(-b)$. Thus, the results do not
depend on the regularization, which demonstrates the validity of the prescription.

\end{appendix}
\end{document}